\documentclass[groupedaddress, superscriptaddress, citesort]{revtex4}
\usepackage{indentfirst}
\usepackage{graphicx}
\usepackage{color}
\usepackage{amsmath,amssymb}
\input{amssym.def}
\def\ra{\rangle}
\def\la{\langle}
\usepackage[colorlinks=true, allcolors=black, citecolor=black, linkcolor=black, urlcolor=black]{hyperref}

\begin{document}
\title{Quantum Recurrence Plot Algorithm Based on Quantum Principal Component Analysis}
\author{Hanhuai Zhu\footnote{E-mail: zhuhanhuai@163.com}}
\affiliation {School of Mathematical Sciences, Capital Normal University, Beijing 100048, China}
\author{Jingjing Huang}
\affiliation {School of Science, Beijing Information Science and Technology University, Beijing 102206, China}
\author{Zhi-Xi Wang\footnote{E-mail: wangzhx@cnu.edu.cn}}
\affiliation{School of Mathematical Sciences, Capital Normal University, Beijing 100048, China}
\author{Shao-Ming Fei\footnote{E-mail: feishm@cnu.edu.cn}}
\affiliation{School of Mathematical Sciences, Capital Normal University, Beijing 100048, China}
\maketitle
\noindent{\bf ABSTRACT}

\noindent
 Recurrence Plot (RP) is a method employed to analyze the periodicity, chaoticity, and nonlinear characteristics of complex systems. Quantum Principal Component Analysis (QPCA), on the other hand, achieves dimensionality reduction of sample data using density matrices based on quantum circuits. We improve the distance threshold function of the recurrence plot algorithm using a density operator conceptually equivalent to the covariance matrix, integrate it with quantum circuits, and thereby develop a Quantum Recurrence Plot (QRP) algorithm. This algorithm achieves ultra-high efficiency in parallel computing, reduces computational costs, and simultaneously upgrades the traditional grayscale recurrence plot to colored heatmaps, enabling a better revelation of the system's dynamical characteristics.



\section{Introduction}

Dynamical systems exhibit geometric structures with complex and self-similar properties. The recurrence plot (RP), as a widely applied fractal theory \cite{ref1}, is not only used to analyze the recurrence of high-dimensional phase space trajectories, but also characterizes the nonlinear dynamic order and unstable periodic orbits \cite{ref2,ref3,ref4}. By mapping sample data into phase space and computing under a threshold function, low-rank matrices are generated to visually interpret the dynamical characteristics.

The recurrence plot has evolved into recurrence quantification analysis (RQA) and cross recurrence plot (CRP) \cite{ref5,ref6}. The former primarily introduced multiple metrics aimed at quantitative analysis of recurrence plot, while the latter was designed to handle the correlation between two time series by constructing recurrence plots. Later, to understand the dynamical characteristics of multidimensional time series, multidimensional recurrence quantification analysis (MdRQA) was proposed \cite{ref7}. In 2019, Sebastian Wallot further developed multidimensional cross recurrence quantification analysis (MdCRQA) based on the CRQA method \cite{ref8}. However, these advancements predominantly involved black-and-white or grayscale images, which were insufficient for revealing dynamical features. Subsequently, as a comprehensive review in the field of recurrence plot, the authors in \cite{ref9} thoroughly discussed and summarized the selection of thresholds by examining the impact of different thresholds and distance metrics on recurrence plot, and proposed an extension from the classical binary-threshold recurrence plot to continuous color-scale recurrence plot. That work also proposes a continuous and differentiable treatment of threshold functions, leading to improved distance threshold functions that output continuous recurrence values. This advancement transformed recurrence plots from traditional ``black-and-white images" into more informative heatmaps, making them more suitable for input into machine learning models.

The quantum computing is receiving increasing attention due to its parallel computing advantages. By combining quantum circuits with classical algorithms \cite{ref10,ref11,ref12,ref13,ref14,ref15,ref16}, many quantum machine learning algorithms have been developed, effectively improving computational efficiency and achieving algorithm acceleration. Among them, QPCA achieves acceleration compared to classical PCA algorithm by solving the eigenvalue problem of $M$-dimensional density matrix \cite{ref17}. Later, based on exchange testing as an alternative to phase estimation, the QPCA \cite{ref18,ref19} was proposed, which is commonly used in low rank density matrix feature extraction scenarios or classification tasks, and has low complexity and is easy to implement.

In response to the above issues, we propose the QRP algorithm based on the QPCA algorithm \cite{ref19,ref20}. Compared to the classical RP, our idea is to use the covariance matrix to improve the distance threshold function. For the quantum principal component analysis method, the covariance matrix is represented by the density matrix, which is the quotient of covariance matrix and trace of covariance matrix. Therefore, in this paper, we use the density matrix to improve the distance threshold function. We calculate the function values and then represent the function values in the density matrix. We use quantum circuits based on exchange test operators to prepare quantum states and perform measurements, and finally generate a low rank heat recurrence plot. The rest of the paper is organized as follows: In the second section, we describe the classic RP algorithm, QPCA algorithm and improved QRP algorithm. In the third section, we conduct experimental verification and discuss the advantages and characteristics of the improved QRP algorithm. Finally, our conclusion is presented in the fourth section.

\section{Methodology}

\subsection{RP algorithm}
The classic recurrence plot is a 2D figure\cite{ref21}. Figure 1 shows two typical recurrence plot. The figure on the top (a) represents a mutation pattern, which is characterized by a large number of white areas and large black block structures. This is due to the rapid changes in the dynamic system, and can be used to test the mutation phenomenon of the system. The figure on the bottom (b) is a periodic pattern. For oscillatory systems, the recurrence plot has a trend in the main diagonal direction and is characterized by a periodic recurrence structure, which is more like a chessboard pattern, with the period being the vertical distance between adjacent diagonal lines. The principle of the method to deal with the problem is to reconstruct the time series ${x_1},{x_2},...,{x_n}$ into a multidimensional phase space through time delay and embedding dimension, in order to study its chaotic and nonlinear characteristics. The reconstructed phase space trajectory obtained is given by
\[
V=
\left(
\begin{array}{lcr}
\begin{aligned}
& V_1  \\
& V_2   \\
 &~\vdots\\
& V_N
\end{aligned}
\end{array}
\right)
=
\left(
\begin{array}{lcr}
\begin{aligned}
&& x_{1}~~~~~~~~&~~~x_{1+\tau}~~~~~~~~\dots&~~~~~~x_{1+(m-1)\tau}~~& \\
&& x_{2}~~~~~~~~&~~~x_{2+\tau}~~~~~~~~\dots&~~~~~~x_{2+(m-1)\tau}~~ &\\
&& ~\vdots~~~~~~~~~~ & ~~~~~~\vdots \quad& ~\vdots~~~~~~~~\\
&& x_{N}~~~~~~~~&~~~x_{N+\tau}~~~~~~~~\dots&~~~~~~x_{N+(m-1)\tau}~~ &\\

\end{aligned}
\end{array}
\right).
\]

Among them, $V_i$ represents the $i$-th state, $N=n-(m-1)\tau$ is the total number of recurrence points, $m\ge1$ is the embedding dimension, and $\tau\ge1$ is the time delay, where the time delay is determined by a nonlinear analysis function mutual information function that can measure the random correlation between two random variables. The calculation of embedding dimension relies on Equations (1) - - (3) proposed by Matthew B. Kennel, which serve as indicators to determine the embedding dimension of time series with attractors.
\begin{eqnarray}\label{E:1.1}
R_m^2(i,n(i,k))=\sum\limits_{k=0}^{m-1}[x(i+k\tau)-x_{n(i,k)}(i+k\tau)]^2,i=1,3,...,n-(m-1)\tau,
\end{eqnarray}

 \begin{eqnarray}\label{E:1.2}
R_{m+1}^2(i,n(i,k))=R_m^2(i,n(i,k))+[x(i+k\tau)-x_{n(i,k)}(i+k\tau)]^2,
\end{eqnarray}
\begin{eqnarray}\label{E:1.3}
\left(\frac{R_{m+1}^2(i,n(i,k))-R_m^2(i,n(i,k))}{R_m^2(i,n(i,k))}\right)^{1/2}>{R_\mathrm{tol}}.
\end{eqnarray}

Among them, $R_m^2(i,n(i,k))$ represents the distance between the reconstructed vector and its nearest neighbor when the embedding dimension is $m$, and $n(i,k)$ is an integer determined by $i$ and $k$. By selecting an appropriate $R_\mathrm{tol}$, the optimal embedding dimension will be obtained when Equation (3) is greater than $R_\mathrm{tol}$ probability of the reconstructed state returning to its neighboring state. This distance threshold function is not fixed. Considering the computational complexity, we usually use the Heaviside function, also known as the $H$ function, which is defined as:
\begin{equation}
R_{ij}(\overline{V})
= H(x)
= H\bigl( \| \overline{V}_i - \overline{V}_j \| - \varepsilon \bigr)
=
\begin{cases}
1, & \| \overline{V}_i - \overline{V}_j \| < \varepsilon \\[6pt]
0, & \text{otherwise}
\end{cases}
\qquad
\overline{V}_i \in \mathbb{R}^m, \quad i,j = 1,2,\ldots,N.
\end{equation}

If $x>0$, $H(x)=1$; if $x<0$, $H(x)=0$, so the recurrence function $R_{ij}$ takes the value 0 or 1, $i$ represents the number of rows, and $j$ represents the number of columns. When $R_{ij}$ is 1, it is represented by black dots in the recurrence plot, which means that the distance between two states is less than the threshold; if $R_{ij}$ is 0, it is represented by a white dot in the recurrence plot, which means that the distance between the two states is greater than the threshold. $\| . \|$ represents the Euclidean norm.

\begin{center}
 \centerline
 {\includegraphics[scale=0.50]{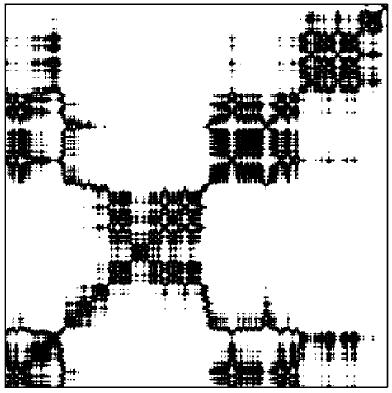}}\vskip3mm
\centering{\small {\bf }\ (a) \label{fig5}}
\end{center}

\begin{center}
 \centerline
 {\includegraphics[scale=0.50]{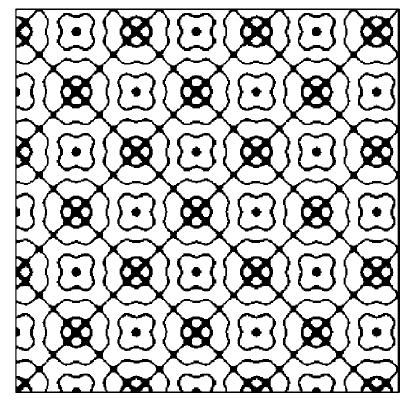}}\vskip3mm
\centering{\small {(b) }\\ {\bf Figure 1}  Typical structural pattern of recurrence plots. \ \label{fig5}}
\end{center}

\subsection{Recurrence Quantification Analysis}
Recurrence quantification analysis (RQA) quantitatively describes a recurrence plot based on the density of recurrence points, the distribution of diagonal structures, and the distribution of vertical structures. In this paper, the following RQA metrics are adopted:
\begin{enumerate}
    \item \textbf{RR (Recurrence Rate)}: Represents the density of recurrence points in the recurrence plot, defined as the ratio of the number of recurrence points to the total number of matrix entries. It is associated with the nonlinearity and nonstationarity of the time series. The basic formula is given by:
    \begin{equation}
    RR = \frac{1}{N} \sum_{i,j=1}^{N} R_{i,j}.
    \end{equation}
    where \( R_{i,j} \) indicates the state where the distance between any two points in the system is less than a predefined threshold.
    \item \textbf{DET (Determinism)}: Represents the ratio of recurrence points forming diagonal lines to the total number of recurrence points in the recurrence plot. This metric is often related to determinism and can distinguish recurrence points that form line segments from isolated recurrence points. The basic formula is:
    \begin{equation}
    DET = \left( \sum_{l=l_{\min}}^{N} l P(l) \right) \cdot \left( \sum_{l=1}^{N} l P(l) \right)^{-1}.
    \end{equation}
    where \( l \) denotes the length of a diagonal structure, \( l_{\min} = 2 \), and \( P(l) \) represents the probability of finding a diagonal structure of length \( l \) in the recurrence plot.
    \item \textbf{L (Average Diagonal Length)}: Represents the average length of diagonal structures, which is equivalent to the average time during which two trajectories remain close to each other. The formula is given by:
    \begin{equation}
        L = \left( \sum_{l=l_{\min}}^{N} l P(l) \right) \cdot \left( \sum_{l=l_{\min}}^{N} P(l) \right)^{-1}.
    \end{equation}
    where \( P(l) \) denotes the probability of finding a diagonal structure of length \( l \) in the recurrence plot.
    \item \textbf{ENTR (Entropy)}: Represents the Shannon entropy when a diagonal structure of length \( l \) is found with probability \( P(l) \), reflecting the complexity of the recurrence plot in terms of its diagonal structures. The basic formula is:
    \begin{equation}
        ENTR = -\sum_{l=l_{\min}}^{N} P(l) \ln P(l).
    \end{equation}
    \begin{equation}
        P(l) = p(l) / N.
    \end{equation}
    where \( p(l) \) is the probability density of the diagonal length distribution, and \( P(l) \) represents the probability of finding a diagonal structure of length \( l \) in the recurrence plot.
\end{enumerate}

\subsection{QPCA algorithm}
Now we describe the basic symbols and methods. Consider a dataset of $N$ training samples denoted by ${x_0,x_1,...,x_{N-1}}$, where $x_i=(x_{0i},x_{1i},...,x_{(M-1)i})^T, i=0,1,...,N-1$ represents one sample and the sample consists of $M$ features. The matrix composed of these $N$  samples is denoted as $X$, i.e

\[
X=
\left(
\begin{array}{lcr}
\begin{aligned}
&& x_{00}~~~~~~~~&~~~x_{01}~~~~~~~~\dots&~~~~~~x_{0(N-1)}~~& \\
&& x_{10}~~~~~~~~&~~~x_{11}~~~~~~~~\dots&~~~~~~x_{1(N-1)}~~ &\\
&& ~\vdots~~~~~~~~~~ & ~~~~~~\vdots \quad& ~\vdots~~~~~~~~\\
&& x_{(M-1)0}~~~~~~~~&~~~x_{(M-1)1}~~~~~~~~\dots&~~~~~~x_{(M-1)(N-1)}~~ &\\

\end{aligned}
\end{array}
\right).
\]
In a matrix, each column represents a sample, and each row represents a centered (mean 0) feature, denoted as $z_i=(x_{i0},x_{i1},...,x_{i(N-1)})$. Therefore, the covariance between any two features is $var(z_i,z_j)=E[(z_i-E(z_i))(z_j-E(z_j))^T]=\sum\limits_{k=0}^{N-1}x_{ik}x_{jk}/N,i,j=0,1,...,M-1$. Based on the number of features, an $MM$-dimensional covariance matrix can be obtained, which we denote as $XX^T$. By establishing a new coordinate system $P(p_0,p_1,...,p_{d-1})$, where $d<M$, the projection of the sample points in the new coordinate system is $P^Tx_i$. The new projection reduces the dimensionality of the data, and the more dispersed the data, the greater the amount of information. Therefore, variance can be used to describe it, and the diagonal elements of the covariance matrix are the variances of each dimension. The covariance matrix after projection is $P^TX(P^TX)^T=P^TXX^TP$, and because the new coordinate system is composed of standard orthogonal bases, the optimization objective can be written as:
\begin{align}
\max \ \mathrm{tr}\left( P^{T} X X^{T} P \right) ,\\
\text{s.t.} \quad P^{T} P = I \nonumber.
\end{align}

By using the Lagrange multiplier method for equation (10), it can be concluded that only by performing eigenvalue decomposition on $XX^T$ and taking the eigenvectors corresponding to the largest d eigenvalues to form a new matrix, the solution of principal component analysis can be obtained. Principal component analysis mainly decomposes the covariance matrix, but in quantum algorithms, density operators are commonly used to represent the matrix. The vector $z_i=(x_{i0},x_{i1},...,x_{i(N-1)})$ composed of individual features from all samples can be represented as $|z_i\ra=\frac{\sum\limits_{k=0}^{N-1}x_{ik}|k\ra}{|z_i|}$ in terms of its quantum state, where $|z_i|=\sqrt{\sum\limits_{k=0}^{N-1}|x_{ik}|^2}$ represents the quantum state of matrix $X$,
\begin{equation}
    |X\rangle = \frac{1}{\sqrt{\beta}} \sum_{i=0}^{M-1} |i\rangle |z_i\rangle,
\end{equation}
among which:

\begin{equation}
    \beta = \sum_{i=0}^{M-1} |z_i|^2\nonumber.
\end{equation}
Equation (11) is a composite system composed of $|i\ra$ and $|z_i\ra$, where $|i\ra$ is the first system and $|z_i\ra$ is the second system. Therefore, the partial trace $\mathrm{tr_2}({|X\ra}{\la X|})$ of operator ${|X\ra}{\la X|}$ is the density operator of the first system, denoted as $\rho$. Simplifying equation $\mathrm{tr_2}(|X\ra \la X|)=\frac{\mathrm{tr_2}((\sum\limits_{k=0}^{M-1}|k\ra|z_k\ra)(\sum\limits_{i=0}^{M-1}\la i|\la z_i|))}{\beta}$ yields the density operator $\rho$, which is exactly the trace of the covariance matrix divided by the covariance matrix

\begin{equation}
   \rho=\mathrm{tr_2}({|X\ra}{\la X|})=\frac{XX^T}{\mathrm{tr}(XX^T)}.
\end{equation}

\subsection{QRP algorithm}
Therefore, in this article, we use a density operator such as formula (12) to improve the threshold function, and use the quantum phase estimation algorithm (QPE) and swap test (ST) operator in the quantum circuits. Through the preparation and measurement of quantum states, we calculate the main eigenvalues and principal components, and further optimize the density matrix to generate a quantum recurrence plot. It should be noted that we do not directly prepare quantum states for the sample, but based on the reconstructed sample $V_1,V_2,...,V_N$ in phase space, rely on the purification procedure for quantum state preparation. The matrix form of the density operator in the above formula is
 \begin{equation}
\rho=
\left(
\begin{array}{lcr}
\begin{aligned}
&& \rho_{00}~~~~~~~~&~~~\rho_{01}~~~~~~~~\dots&~~~~~~\rho_{0(M-1)}~~& \\
&& \rho_{10}~~~~~~~~&~~~\rho_{11}~~~~~~~~\dots&~~~~~~\rho_{1(M-1)}~~ &\\
&& ~\vdots~~~~~~~~~~ & ~~~~~~\vdots \quad& ~\vdots~~~~~~~~\\
&& \rho_{(M-1)0}~~~~~~~~&~~~\rho_{(M-1)1}~~~~~~~~\dots&~~~~~~\rho_{(M-1)(M-1)}~~ &\\

\end{aligned}
\end{array}
\right).
\end{equation}

To compute the eigenvalues of the density matrix and generate the quantum recurrence plot (QRP), we provide two solution methods for high-dimensional features ($m > 2$) and low-dimensional features ($m = 2$), respectively. For low-dimensional feature samples, to prepare the quantum state, we represent the density operator shown in Equation (13) in the form of spectral decomposition $\rho=\sum\limits_{i=1}^{M}\lambda_i|p_i\ra \la p_i|$ where $\lambda_i$ is the eigenvalue of $\rho$ and $|p_i\ra$ is the corresponding eigenvector. Then, according to the purified definition, we represent the density operator $\rho$ in the following form: $\rho=\mathrm{tr_2}(|\phi\ra \la \phi|)$, $\phi=\sum\limits_{i=1}^{M}\sqrt\lambda_i|p_i\ra|p_i\ra$. Then, according to reference\cite{ref22}, the trace of the square of $\rho$ is exactly equal to the expectation value $\mathrm{tr}(\rho^2)=\la \phi|\la \phi|ST|\phi\ra|\phi\ra$ of the swapping operator in system $|\phi\ra|\phi\ra$. Therefore, the quantum circuit diagram we designed is shown in Figure 2: the ST operator is mainly used to swap the quantum state of the first system in $|\phi\ra|\phi\ra$, i.e. $ST(\sum\limits_{i}^{}\sum\limits_{i'}^{}\sqrt{\lambda_i}\sqrt{\lambda_{i'}}|p_i\ra|p_i\ra|p_{i'}\ra|p_{i'}\ra)=\sum\limits_{i}^{}\sum\limits_{i'}^{}\sqrt{\lambda_i}\sqrt{\lambda_{i'}}|p_{i'}\ra|p_i\ra|p_{i}\ra|p_{i'}\ra$.

\begin{center}
 \centerline
 {\includegraphics[scale=0.50]{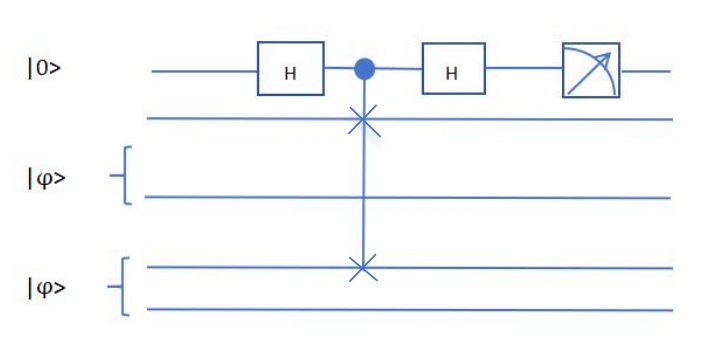}}\vskip3mm
\centering{{\bf Figure 2}  Quantum circuit diagram for computing the square of density operator using ST operator. \ \label{fig5}}
\end{center}
\begin{center}
 \centerline
 {\includegraphics[scale=0.50]{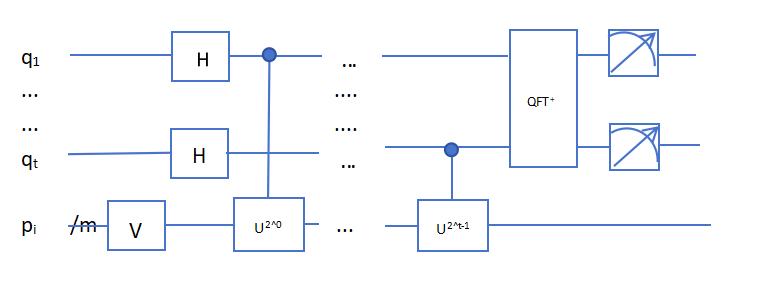}}\vskip3mm
\centering{{\bf Figure 3}  Quantum circuit diagram for computing eigenvalues of a density operator using quantum phase estimation. \ \label{fig5}}
\end{center}

Using the above quantum circuit for measurement, the probabilities of obtaining $|0\ra$ and $|1\ra$ are $P(0)=(1+\la \phi|\la \phi|ST|\phi\ra|\phi\ra)/2$ and $P(1)=(1-\la\phi|\la\phi|ST|\phi\ra|\phi\ra)/2$, respectively. Finally, the trace of the square of the operator is equal to $P(0)-P(1)$.  Then, we can use the following formula $\lambda_1+\lambda_2=\mathrm{tr}(XX^T),{\lambda_1}^2+{\lambda_2}^2={\mathrm{tr}(XX^T)}^2\mathrm{tr}(\rho^2)$
 to calculate the main feature values $\lambda_1=\mathrm{tr}(XX^T)\frac{1+\sqrt{1-2(1-\mathrm{tr}(\rho^2))}}{2}$, $\lambda_2=\mathrm{tr}(XX^T)\frac{1-\sqrt{1-2(1-\mathrm{tr}(\rho^2))}}{2}$.

For high-dimensional feature samples, the quantum phase estimation algorithm (QPE) is employed to solve for the eigenvalues. The corresponding quantum circuit is shown in Figure 3. First, the quantum state is prepared using a unitary transformation $V$, and a superposition state is constructed by applying $t$ Hadamard gates to the first register:\[
|\psi_1\rangle = \frac{1}{2^{t/2}} \sum_{k=0}^{2^{t-1}-1} |k\rangle |u\rangle.
\]Controlled by the $j$-th qubit ($j = 0, 1, \dots, t-1$) in the first register, the quantum state of the $j$-th $U$ is applied to the second register, transferring the phase into the amplitudes of the first register. The quantum state thus evolves into:
\[
|\psi_2\rangle=\frac{1}{2^{t/2}} \sum_{k_1=0}^{1} \cdots \sum_{k_t=0}^{1} |k_1 \cdots k_t\rangle \otimes U^{k_1 2^{t-1} + k_2 2^{t-2} + \cdots + k_t 2^0} |u\rangle.
\]Since QPE is based on $|u\rangle = e^{2\pi i \varphi}|u\rangle$, i.e., $e^{2\pi i \varphi}$ is the eigenvalue of $U$ and $|u\rangle$ is the corresponding eigenvector, the above expression can be rewritten as:\[
|\psi_3\rangle=\frac{1}{2^{t/2}} \sum_{l=0}^{t} \left( |0\rangle + e^{2\pi i \varphi 2^{t-l}} |1\rangle \right) |u\rangle.
\]
Then, the inverse quantum Fourier transform ($\text{QFT}^\dagger$) is applied to the first register to transfer the phase stored in the amplitudes into the basis states:\[
\text{QFT}^\dagger \left[ \frac{1}{2^{t/2}} \sum_{l=0}^{t} \left( |0\rangle + e^{2\pi i \varphi 2^{t-l}} |1\rangle \right) \right] = |\varphi_1 \cdots \varphi_t\rangle.
\]Finally, the eigenvalue $\varphi$ can be obtained through a simple binary transformation.

The process of the algorithm is completed in this way. For the new sample, the above operations are performed to obtain an improved quantum recurrence plot. The difference between QRP and RP algorithms is shown in Figure 4: where $t_1,t_2,...,t_n$ is the sample data, and after calculating the embedding dimension and time delay, the reconstructed sample data $V_i,i=1,2,...m$ in phase space is obtained. The classical recurrence plot calculates the distance threshold function to directly generate the recurrence plots, while the quantum recurrence plot uses the quantum form of the covariance matrix to calculate the reconstructed state using the covariance formula. Considering the computational cost of large data, we then optimized the improved quantum recurrence plot using quantum circuits based on exchange testing, resulting in our improved quantum recurrence plot.The classical recurrence plot (RP) is generated by directly computing the distance threshold function, producing the recurrence image. In contrast, the quantum recurrence plot (QRP) employs a quantum formulation of the covariance matrix to process the reconstructed states. Subsequently, a quantum circuit based on the swap test and quantum phase estimation (QPE) is used to optimize the improved recurrence plot, yielding the quantum recurrence plot.

\begin{center}
 \centerline
 {\includegraphics[scale=0.50]{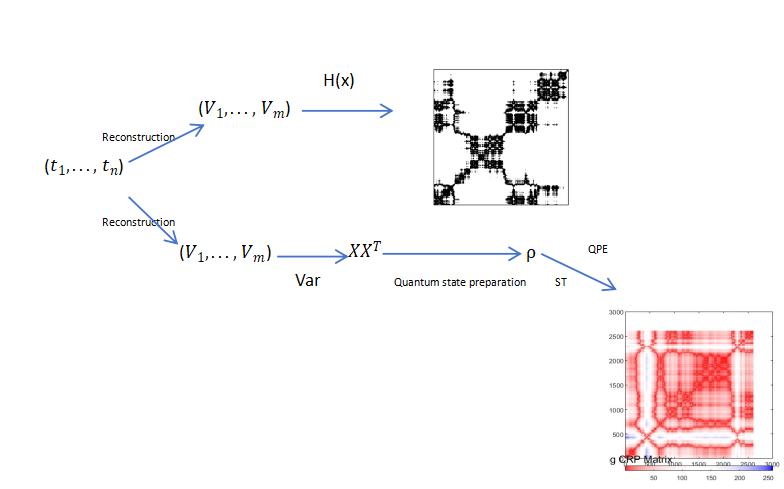}}\vskip3mm
\centering{{\bf Figure 4}  Algorithm flow of classical recurrence plots and quantum recurrence plots. \ \label{fig5}}
\end{center}

\section{EXPERIMENTS}

\subsection{Recurrence Plot Comparison and Analysis}
The data used in this study are historical trading data of individual stocks obtained from NetEase Finance. Specifically, we selected the daily closing price time series of the SSE 50 Index from January 5, 2005 to May 10, 2021, resulting in a total of 3970 data points. This dataset is relatively large and exhibits complex dynamical characteristics, making it suitable for validating the effectiveness of the proposed method.

First, the false nearest neighbor (FNN) method was employed to determine the embedding dimension of the time series. As shown in Figure 5, the embedding dimension stabilizes after reaching 6, which is therefore selected as the optimal dimension.
Subsequently, phase space reconstruction was performed. The classical recurrence plot (RP) was first constructed by setting the threshold $\varepsilon = 0.1$ and computing the distance threshold function, yielding a low-rank matrix. The resulting RP is presented in Figure 6a.
For the quantum recurrence plot (QRP), the initial steps are identical to those of the classical RP. After obtaining the phase space states, the covariance matrix is computed. A quantum circuit centered on quantum phase estimation (QPE) is then employed to prepare and process the classical data. Here, the quantum state is initialized from $|0\rangle$ to a purified form. Finally, the QRP is obtained, as shown in Figure 6b.

\begin{center}
 \centerline
 {\includegraphics[scale=0.50]{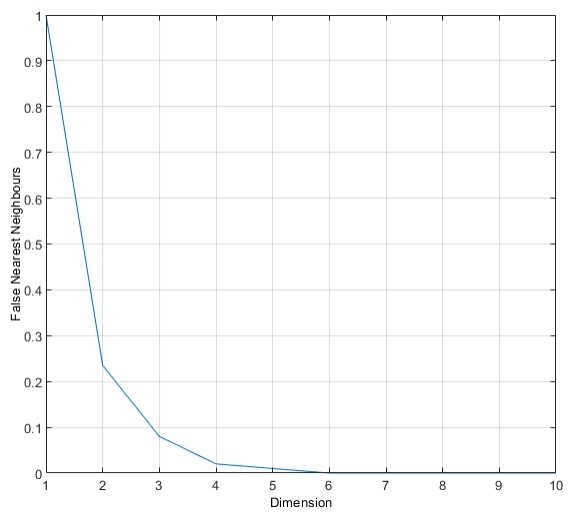}}\vskip3mm
\centering{{\bf Figure 5}  Embedding dimension analysis. }\ \label{fig5}
\end{center}

\begin{center}
 \centerline
 {\includegraphics[scale=0.60]{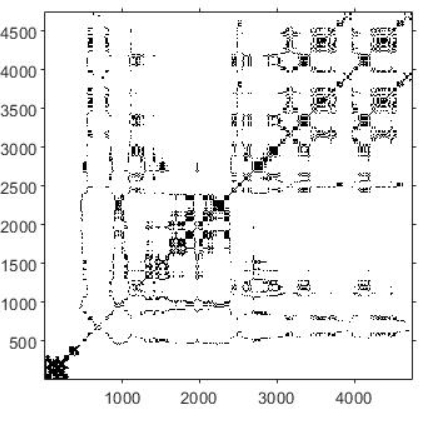}}\vskip3mm
\centering{\small {\bf }\ (a) \label{fig5}}
\end{center}
\begin{center}
 \centerline
 {\includegraphics[scale=0.75]{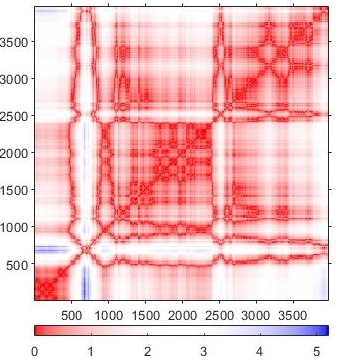}}\vskip3mm
\centering{\small {(b) }\\ {\bf Figure 6}  Comparison of recurrence plots. (a) Classical recurrence plot; (b) Quantum recurrence plot based on density matrix and quantum phase estimation.} \ \label{fig5}
\end{center}

\subsection{Recurrence Quantification Analysis}

To further analyze the dynamical characteristics of the recurrence plot, recurrence quantification analysis (RQA) was performed on both the classical recurrence plot (RP) and the quantum recurrence plot (QRP). The results are presented in Figure 7. By comparing Figure 7 a,b, we observe that the QRP exhibits clearly visible periodic diagonal lines that appear periodically along the main diagonal, indicating that the system possesses strong periodic behavior or quasi-periodic oscillations. In contrast, the RP shows relatively sparse recurrence points and weaker representational capability.Observing the density of recurrence points near the main diagonal, the RP exhibits lower density due to the presence of noise or parameter drift. In comparison, the QRP exhibits higher recurrence point density, and thanks to reduced noise interference, its $DET$ is higher.
Furthermore, we find that the QRP has longer diagonal lines, which compensates for the deficiency in the RP caused by parameter drift or other factors that reduce the frequency of state revisit in phase space. Additionally, due to the improved threshold function, the QRP more accurately reflects the distances between system states in phase space. Consequently, no large bright blocks or white band-like regions appear in the QRP, indicating that the system does not undergo large-scale state transitions. Owing to the limitations of its threshold function, the RP struggles to form long diagonal structures, which accounts for its lower entropy ($ENTR$) compared to the QRP.

\begin{center}
 \centerline
 {\includegraphics[scale=0.40]{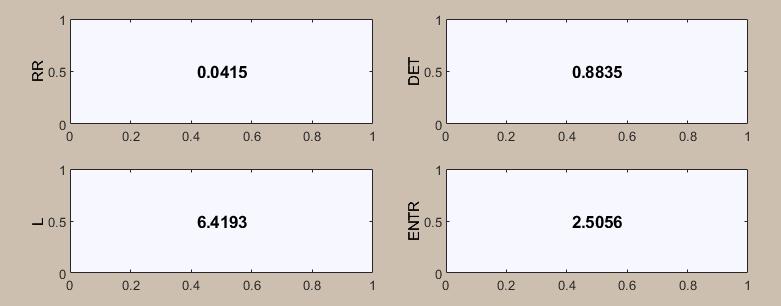}}\vskip3mm
\centering{\small {\bf }\ (a) \label{fig5}}
\end{center}
\begin{center}
 \centerline
 {\includegraphics[scale=0.40]{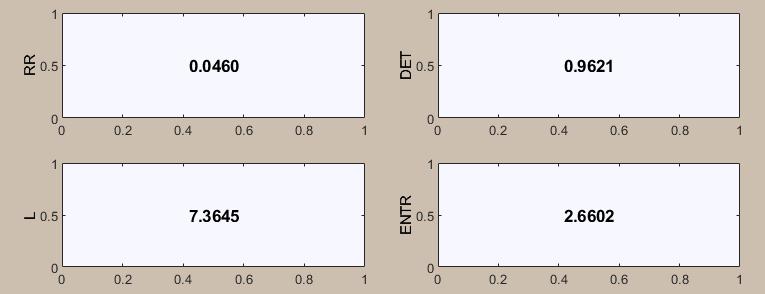}}\vskip3mm
\centering{\small {(b) }\\ {\bf Figure 7}  Recurrence plot comparison: (a) Classical recurrence plot; (b) Quantum recurrence plot based on covariance matrix and quantum phase estimation.} \ \label{fig5}
\end{center}

\subsection{ Complexity Analysis}
To further verify the computational efficiency of the two methods, this section compares the time complexity of the classical recurrence plot (RP) with that of the proposed quantum circuit-based QRP. The results are presented in Figure 8.
The computation of a classical recurrence matrix requires calculating the Euclidean distances (or other norms) between all pairs of states, followed by comparisons with a threshold under the threshold function. Consequently, this step must traverse \(N \times N\) states, yielding a time complexity of \(O(m^{2})\). In contrast, the quantum circuit based on quantum phase estimation (QPE) enables parallel computation across all quantum bits. The preparation of quantum states involves a constant number of operations, which can be neglected. The resulting complexity is \( O(\log m)\) quantum gate operations.
Therefore, the QRP employing density matrix encoding and QPE achieves an exponential speedup in theory compared to the classical RP. This advantage stems from quantum parallelism and the logarithmic query complexity of QPE, offering substantial potential for processing long-sequence data such as financial time series, electroencephalography (EEG) data, and meteorological data. It should be noted, however, that the aforementioned speedup relies on sufficiently long qubit coherence times and ideal quantum hardware. On noisy intermediate-scale quantum (NISQ) devices, a trade-off between computational overhead and error mitigation must be carefully considered.

\begin{center}
 \centerline
 {\includegraphics[scale=0.35]{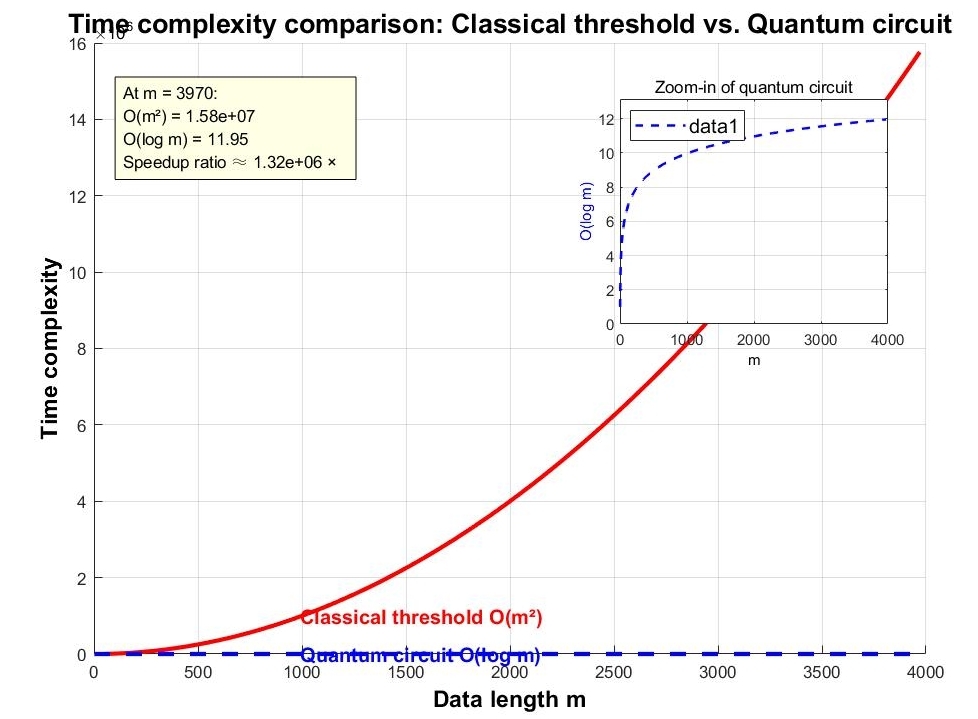}}\vskip3mm
\centering{{\bf Figure 8}  Time complexity comparison: Classical recurrence plot ($O(m^2)$) vs. Quantum recurrence plot ($O(log m)$)} \ \label{fig5}
\end{center}

\section{CONCLUSIONS}

In this paper, the classical recurrence plot (RP) is improved in two aspects. First, a quantum circuit is employed to prepare the density matrix and modify the threshold function, generating a colored QRP that more intuitively reflects the dynamical characteristics of the system. Second, eigenvalue solving based on quantum phase estimation and the swap test operator is implemented, achieving both noise reduction and dimensionality reduction of the data.Specifically, the parallelism of quantum circuits significantly accelerates the construction of recurrence plots and reduces time overhead. Quantum principal component analysis (qPCA) is applied to reduce the dimensionality of the recurrence plot, preserving the main information of the original data while decreasing computational cost. Compared to classical RPs, the QRP employs a color-based distance metric to reflect the trajectory behavior of the reconstructed phase space, overcoming the limited information capacity of grayscale images and revealing the nonlinear dynamical features of the system more clearly.Experimental results demonstrate that the QRP outperforms the classical RP in terms of recurrence rate, determinism, diagonal structure integrity, and complexity characterization, validating the effective integration of classical and quantum algorithms.
Nevertheless, the estimation of time delay and embedding dimension currently still relies on classical computational methods. Future work may explore transferring these preprocessing steps to quantum circuits as well, achieving a fully quantum pipeline from data input to recurrence plot generation. We anticipate that this improvement will further enhance computational efficiency, particularly for the analysis of long-sequence, high-dimensional time series.
In summary, the combination of quantum circuits based on parallel computing with classical machine learning methods offers an efficient and promising new paradigm for nonlinear time series analysis.

\section{Acknowledgments}
This work is supported by the National Natural Science Foundation of China (NSFC) under Grant Nos. 12171044; the specific research fund of the Innovation Platform for Academicians of Hainan Province.

\end{document}